\documentclass[aps,prl,showpacs,amsmath,amssymb,twocolumn,superscriptaddress]{revtex4-1}
\usepackage{graphicx}% Include figure files
\usepackage{epsfig}
\usepackage{color}
\usepackage{amsmath}

\newcommand{\dudu}{\langle u_{dc}'|u_{dc}'\rangle}

\newcommand{\dcndcm}{\langle \chi_n'|\chi_m'\rangle}
\newcommand{\ddcnddcm}{\langle \chi_n''|\chi_m''\rangle}

\newcommand{\dudcn}{\langle u_{dc}'|\chi_n'\rangle}
\newcommand{\dudcm}{\langle u_{dc}'|\chi_m'\rangle}
\newcommand{\cncm}{\langle \chi_n|\chi_m\rangle}

\newcommand{\dudco}{\langle u_{dc}'|\chi_o'\rangle}
\newcommand{\dcodcp}{\langle \chi_o'|\chi_p'\rangle}

\newcommand{\Tt}{\tau}

\begin{document}

\title{Tension-induced non-linearities of flexural modes in nanomechanical resonators}

\author{Rapha\"el Khan}
\email[]{raphael.khan@aalto.fi}
\author{F. Massel}
\author{T.~T.~Heikkil\"a}
\affiliation{Low Temperature Laboratory, Aalto University, P.O. Box 15100, FI-00076 AALTO, Finland}

\newcommand{\tmpnote}[1]%
   {\begingroup{\it (FIXME: #1)}\endgroup}
   \newcommand{\comment}[1]%
       {\marginpar{\tiny C: #1}}
%%% Uncomment the following lines to exclude temporary comments
%\renewcommand{\tmpnote}[1]{}
%\renewcommand{\comment}[1]{}

\date{\today}

\begin{abstract}
We consider the tension-induced non-linearities of mechanical resonators, and derive the Hamiltonian of the flexural modes up to the fourth order in the position operators. This tension can be controlled by a nearby gate voltage. We focus on systems which allow large deformations $u(x)\gg h$ compared to the thickness $h$ of the resonator and show that in this case the third-order coupling can become non-zero due to the induced dc deformation and offers the possibility to realize radiation-pressure-type equations of motion encountered in optomechanics. The fourth-order coupling is relevant especially for relatively low voltages. It can be detected by accessing the Duffing regime, and by measuring frequency shifts due to mode-mode coupling.
\end{abstract}

\maketitle

Recent progress in fabricating nanomechanical resonators has shown how these systems can be used for
 ultrasensitive measurements of mass, force 
 and charge \cite{jensenk._atomic-resolution_2008,chaste_nanomechanical_2012,gavartin_hybrid_2012,bunch_electromechanical_2007}. Within the couple of past years these systems have also entered 
the quantum realm \cite{oconnell_quantum_2010} as superpositions of vibration states and zero-point vibrations 
have been  measured. Even though such measurements can be performed in a regime
where the elastic properties of the resonators could essentially be
considered as linear, the extension to non-linear conditions is well
within reach of the current experimental techniques.

In this paper we consider the generic non-linearities of the
resonators, how these show up in measurements, and how they arise when
the resonators are manipulated electronically. In general, the effect
of non-linearites is twofold: on one hand they modify in an
amplitude-dependent way the resonant frequency of a given normal mode
(Duffing self-non-linearity); on the other, they introduce a coupling
between normal modes. Such non-linearities show up in the presence of
strong external driving, which allows to control the coupling of
different modes or to detect their occupation numbers.  

\begin{figure}
 \includegraphics[scale=0.95]{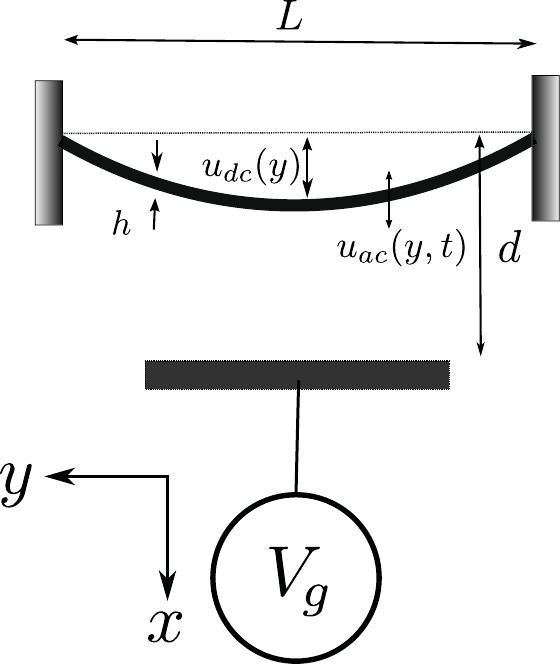}
\caption{Schematic picture of the studied setup :  a metallic beam whose deformation is controlled by a gate  voltage $V_g$ coupled to the beam via capacitance $C(d-u(x,t))$.}
\label{fig:setup}
\end{figure}

Motivated by
the recent advances in fabricating graphene and carbon nanotube
resonators \cite{bunch_electromechanical_2007,huttel:2547}, we
concentrate especially on the regime of thin resonators where the
mechanical deformation can be large compared to the resonator
thickness. In this case, the major source of non-linearity is the
tension induced by the deformation itself. Starting from the
mechanical energy of the deformations, we derive the generic
Hamiltonian of the flexural modes, including non-linearities up to the
fourth order in the vibration amplitudes. In contrast to the results discussed in  Refs.~\cite{PhysRevLett.105.117205,1204.4487},where it is not taken into account, we explicitly consider the 
 dc deformation of the resonator. This additional aspect creates an asymmetry in our system which leads to a cubic non-linearity. The dc deformation, dictating the strength of the non-linearity,  is driven by a
nearby gate voltage as in Fig.~\ref{fig:setup}. Concentrating first on
the Duffing self-non-linearity of the modes, this then allows us to
derive the voltage dependence of the Duffing constant and show that it
changes sign for a certain value of voltage that depends on mode index
and the amount of initial tension. This sign change results primarily from the cubic non-linearity. Therefore, studies of the Duffing
constant reveal information about the parameters of the system, in
particular on the initial tension, which may otherwise be difficult to
obtain by only concentrating on the voltage dependence of the mode
eigenfrequencies. We go on to analyze the inter-mode coupling and show
that the non-linearities allow creating a radiation-pressure-type
coupling between the different flexural modes. Such a coupling allows
realizing optomechanics-type experiments, where one of the modes is
cooled or heated by driving another mode. We provide quantitative
predictions for the optical spring effect (driving-induced frequency
shift) and changes in effective mode damping responsible for the
cooling/heating behavior and show how these can be tuned by the dc
gate voltage.
The general Hamiltonian describing a non-linear resonator is of the form
\begin{multline}
H=\sum_n  \omega_n \hat a^\dagger_n \hat a_n + \sum_{nml} \mathbb{T}_{nml} \hat x_n \hat x_m \hat x_l\\
 + \sum_{nmlk}\mathbb{F}_{nmlk} \hat x_n \hat x_m \hat x_l \hat x_k+{O}(\hat x_o\hat x_n \hat x_m \hat x_l \hat x_k).
\label{eq:generalhamiltonian}
\end{multline}
Here $\hat x_n = a_n^\dagger + a_n$ are the dimensionless position operators. The non-linearities are
 described by the coefficients $\mathbb{T}_{nml}$ and $\mathbb{F}_{nmlk}$. The presence of $\mathbb{T}_{nml}$, like any odd non-linearities,  arises from  an asymmetry in the system. In the following we consider a  mechanical
 resonator exhibiting the non-linearities discussed above. 
We analyse a beam of mass $m$  with length $L$, thickness $h$ and cross-section $S$ suspended on top of a gate capacitor at voltage $V_g$ (see Fig.~\ref{fig:setup}). The flexural vibrations are characterized by the deformation $u(x,t)$ of the beam. Defining $z=y/L$, $u'=\partial_z u$ and introducing the notation $\langle u|v\rangle=\int^1_0 u(z,t)\cdot v(z,t)\; dz$ one can obtain \eqref{eq:generalhamiltonian} from the elastic energy of a resonator \cite{landau_theory_1986}
\begin{eqnarray}\label{fullenergy}
  \varepsilon[u(x,t)]&=&\underbrace{\frac{EI_y}{2L^3}}_{\frac{1}{2}m\omega_0^2}\langle u''|u''\rangle+\underbrace{\frac{T_0}{2L}}_{\frac{1}{2}\tau_0 m\omega_s^2}\langle u'|u'\rangle  \nonumber \\
&+&\underbrace{\frac{ES}{8L^3}}_{\frac{1}{2h^2}m\omega_s^2}\langle u'|u'\rangle^2+\varepsilon_{\rm{force}}[u(x,t)],
\end{eqnarray} 
with  $E$ the Young modulus, $I_y$ the bending moment,
$T_0$ the initial tension of the resonator and $\varepsilon_{\rm{force}}$ the potential energy of the force acting on the resonator. The latter is  
of the form $\varepsilon_{\rm{force}}=-(V_g^2/2)\int^1_0C[d-u(z,t)]dz$, where $C[d]$ is the capacitance between the gate and the beam at the distance $d$. 
In order to arrive at a Hamiltonian of the form given in Eq.~\eqref{eq:generalhamiltonian}, we assume that the gate voltage $V_g$ is the sum of a DC part $V_{dc}$ and a small AC part $V_{ac}\ll V_{dc}$. These voltages  lead to  a static deformation $u_{dc}(x)\ll d$ and a time varying part $u_{ac}(x,t)\ll u_{dc}(x)$. Expanding $u_{ac}(x,t)$  on  an arbitrary basis $\chi_n(x)$, $u_{ac}=h \sum_n y^n(t)\chi_n(x)$, we can write the potential energy containing terms with two, three and four $y^n$s. Writing the Hamiltonian in terms of the stress energy $E_{\rm stress}=m\omega_s^2 h^2/2=ESh^4/(8 L^3)$, these terms are characterized by the dimensionless parameters which we denote by $[\Omega^2]^n_m$, $\Lambda^n_{mo}$, and $\Theta^{no}_{mp}$ for the second-, third- and fourth-order terms, respectively. These parameters are described in detail in the Appendix \cite{supplement}. In particular, they depend on the dc bias voltage $V_{dc}$ and the total tension $T$ in the beam. As discussed below, the latter also depends on $V_{dc}$. The behavior of the coefficient $[\Omega^2]^n_m$ determines the voltage dependence of the eigenfrequency as described in \cite{PhysRevB.67.235414,PhysRevB.84.195433}. For $V_{dc}=0$,  third-order terms $\Lambda^n_{mo}$ vanish because of symmetry ($u_\mathrm{dc}=0$), but for large $V_{dc}$ they grow as $\Lambda^n_{mo} \propto V^{2/3}$. The voltage dependence of the fourth-order terms $\Theta^{no}_{mp}$ on the other hand is weak and in our analytical approximations \cite{supplement} disregarded altogether. 

We arrive at the desired form \eqref{eq:generalhamiltonian} by writing the Hamiltonian in a basis which diagonalises $\left[\Omega^2 \right]^n_m $ and scaling the amplitude $h y_n$ of each mode by its zero point motion $x_{zp_n}=\sqrt{\frac{1}{m\omega_n}}$ \cite{quantumnote}. Non-linearities  are coming primarily from the induced tension, which is maximized for large deformations $u(x)\gg h$. Therefore we concentrate on  systems which allow large deformations, i.e., systems with $d\gg h$. We truncate the expansion to the fourth order as in the model employed above the higher-order couplings are relevant only close to the point where the beam pulls into contact with the gate plane \cite{PhysRevB.84.195433}. 

As an example, a single-layer graphene sheet with $h=0.34$ nm, $L=1$ $\mu$m, $S/h=1$ $\mu$m, $E=1$ TPa and mass density $\rho=1400$ kg/m$^3$ would have $\omega_s=250$ MHz and $E_{\rm stress} = 0.01$ eV. 

\begin{figure}
\includegraphics[width=\columnwidth]{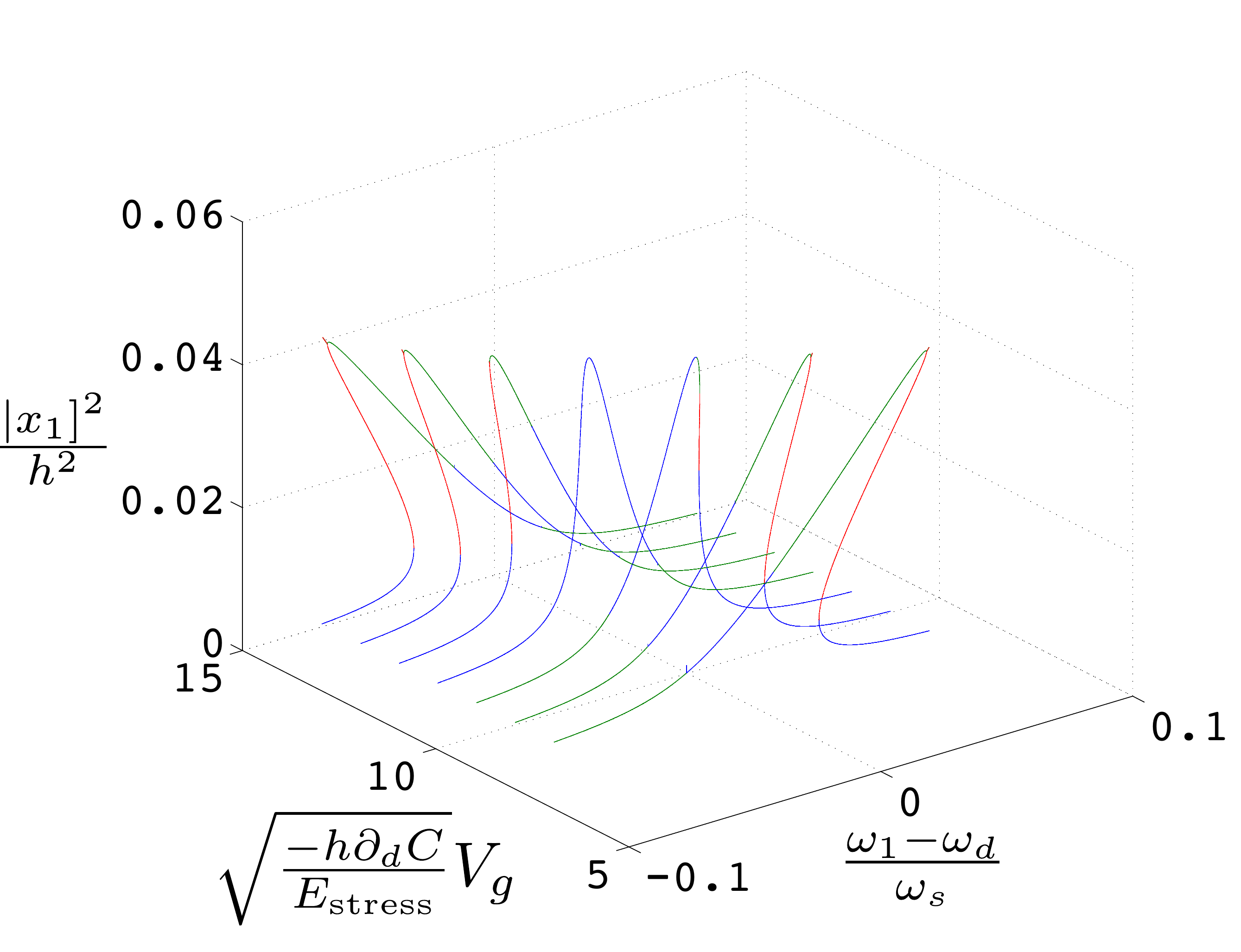}
\caption{(Color online) Frequency response function for the first mode with  $\gamma^2=0.01\, \omega_s^2$ and $|f_d|^2=0.002\, m^2 \omega_s^2 h^2$. Here $E_\mathrm{stress}=ESh^4/(8L^3)$.}
\label{fig:selfamplitude}
\end{figure}
% \begin{eqnarray}
% \Omega^n_m&=&\left(\frac{\omega_0}{\omega_s}\right)^2(n\pi)^4\delta_{n,m}+\left(\frac{\omega_T}{\omega_s}\right)^2(n\pi)^2\delta_{n,m}\nonumber  \\
% &+&\left(\frac{\tilde{V}}{2}\right)^4\frac{1}{T^2}\left[\frac{2}{3}(n\pi^2)\delta_{n,m}+\frac{32}{n m \pi}P_{n}P_m\right]\label{OMEGA} \\
% \Lambda_{nmo}&=&\left(\frac{x_0}{d}\right)\left[2\sqrt{2}n^2\pi \left(\frac{\tilde{V}}{2}\right)^2\frac{1}{T}\frac{P_o}{o}\right]\\
% \Theta^{no}_{mp}&=&\left(\frac{x_0}{d}\right)^2\bigg[(np\pi^2)^2\delta_{p,o}\delta_{n,m}\bigg]
% % \end{eqnarray}

{\it Self-non-linearity.} Let us first consider  the non-linear effects which occur when driving mode $n$ with a driving force $f_d\cos(\omega_d t)$, $\omega_d \approx \omega_n$, disregarding the coupling to the other modes, as in the absence of direct driving of the other modes these would show up only in a higher order in the non-linear coupling constants. We also exclude the special case when 2 or 3 times the mode frequency  matches one of the other mode frequencies \cite{nayfeh_nonlinear_2008,antonio_frequency_2012,PhysRevLett.109.025503}. Including dissipation the equation of motion for the amplitude $x_n$ of  mode $n$ is
\begin{equation}\label{solomode}
\ddot x_n +\omega_n^2 x_n +\gamma \dot{x}_n + 3 \mathcal{T}_n x_n^2 + 4 \mathcal{F}_n x_n^3 = \frac{f_d}{m}\cos(\omega_d t).
\end{equation}
Here $x_n=x_{zp_n}\hat x_n$, $\omega_n=\omega_s (\Omega_n^n)_d$, $\mathcal{T}_n \equiv \frac{1}{2}\frac{\omega_s^2}{h}\left(\Lambda^n_{nn}\right)_d$, $\mathcal{F}_n\equiv \frac{1}{2}\frac{\omega_s^2}{h^2}\left(\Theta^{nn}_{nn}\right)_d$ and the subscript $d$ denotes that the tensors are written in the basis which diagonalises $[\Omega^2]$. The frequency response equation of \eqref{solomode} can be solved from \cite{nayfeh_nonlinear_2008,kozinsky:253101}
\begin{equation}
 |x_n|^2=\frac{|f_d|^2/m^2}{\gamma^2+\left[(\omega_n-\omega_d)-\frac{3}{8}\frac{D }{\omega_n}|x_n|^2\right]^2},
\end{equation}
where $\gamma$ is the damping constant and $D=4\mathcal{F}_n-\frac{10}{8}\left(\frac{\mathcal{T}_n }{\omega_n}\right)^2= \omega_s^2/(2h^2) \left[(\Theta_{nn}^{nn})_d-5(\Lambda_{nn}^n)_d^2/[2(\Omega_n^n)_d^2]\right]$. This frequency response function is the same as one would get when considering only the fourth-order non-linearity $\mathcal{F}_n$, i.e, a Duffing oscillator. The effect of the third-order non-linearity $\mathcal{T}_n$ is to shift the value of the cubic non-linearity in the frequency response function \cite{kozinsky:253101}.

An example response function obtained for different dc gate voltages is plotted in Fig.~\ref{fig:selfamplitude} and shows up a crossover from $D>0$ (hardening) to $D<0$ (softening). The dc voltage dependence of $D$ is plotted in Supplementary information \cite{supplement} for a few different magnitudes of initial tension $T_0$, characterized by the dimensionless quantity $\tau_0 =4T_0L^2/(E S h^2)=h^2 T_0/(2 E_{\rm stress} L)$. Here we characterize its overall behaviour. The behavior of the Duffing constant depends on the total tension $T$ of the beam which is the sum of the initial tension $T_0$ and the tension induced by the deformation $u_{dc}(x)$ caused by $V_{dc}$. The latter has to be calculated self-consistently from the Euler--Bernoulli equation as discussed in \cite{PhysRevB.67.235414}. In what follows, we describe this behavior in terms of the dimensionless quantity $\tau=4 T L^2/(E S h^2)=h^2 T/(2 E_{\rm stress} L)$. In the limit $d \gg h$ it satisfies \cite{supplement}
\begin{equation} 
\begin{split}
\tau&=\tau_0+2\int_0^1u_{dc}'^2dz\\
 &=\tau_0+\frac{\tilde{V}^4}{96\tau^2}\left(1-\frac{3\sqrt{3}}{\sqrt{\tau}}+\frac{8}{\tau}\right),
\end{split}
\label{eq:tensionequation}
\end{equation}
with $\tilde{V}^2=(-h \partial_d C V_{g}^2)/E_{\rm stress}$. This equation is valid provided the resultant $\tau \gtrsim 1$. In the same limit, we find for the Duffing coefficient of the fundamental mode $n=1$
\begin{equation}
D=\frac{\omega_s^2}{2h^2}\left[\pi^4-\frac{5}{8}\frac{\pi^4 f_1(\tau)^2\frac{\tilde V^4}{\tau^2}}{\frac{4I_y \pi^4}{Sh^2}+\tau\pi^2+f_1(\tau)^2\frac{\tilde V^4}{8\tau^2}}\right],
\label{eq:duffingequation}
\end{equation}
where $f_n(\tau)=2n\pi[2/(n\pi)^2-\sqrt{3\tau}/((n\pi)^2 + 3\tau)]$ for $n$ odd and zero otherwise. For a rectangular beam $I_y/(S h^2)=1/12$, but the overall behavior of $D$ does not greatly depend on the exact shape of the beam. 

Solving Eqs.~\eqref{eq:tensionequation} and \eqref{eq:duffingequation} allows us to find the approximate behavior of the Duffing constant as the gate voltage is tuned. We find that $D$ changes sign at a value of the gate voltage $\tilde V^*$ that can be quite well fitted to the function $\tilde V^* \approx \sqrt{2}\tau^{3/4}+8$ (see \cite{supplement}) or
\begin{equation}
V_{g}^* \approx \frac{h^3}{2.4 (-h\partial_d C)^{1/2} E_{\rm stress}^{1/4}} T^{3/4} +8 \frac{E_{\rm stress}^{1/2}}{-h\partial_d C}.
\end{equation}
Moreover, at large values of the voltage $V_g \gg V_g^*$, $D$ (approximatively) saturates to the value $D_{\rm sat} \approx -120 \omega_s^2/h^2$. 

Contrary to the fundamental mode $n=1$, the deformation induced changes in the Duffing constant of higher-order modes are rather small compared to its value for $V_g=0$.

{\it Non-linear mode coupling.} 
Let us now concentrate on the non-linear coupling between the modes \cite{PhysRevLett.105.117205}. Unlike  in Ref.~\cite{mahboob_phonon-cavity_2012}, where the coupling  between the modes is a time-dependent linear coupling, in our system the introduction of the dc deformation
leads to a radiation-pressure coupling (second term in \eqref{eq:generalhamiltonian}).   The
  regime investigated here is formally analogous to the setup encountered in optomechanical systems
  \cite{marquardt_quantum_2007,PhysRevA.77.033804,Teufel:2011jga,massel_microwave_2011}, where an external driving
  electromagnetic field, coupled to a resonant cavity, alters the characteristic response parameters of a mechanical
  resonator. More specifically, by aptly tuning the pump frequency, it is possible to alter the resonant frequency of
  the mechanical resonator (optical spring effect \cite{PhysRevLett.101.197203,Rocheleau:2010jd,massel_microwave_2011}) and its
  damping, thereby inducing cooling \cite{Teufel:2011jga} or amplification \cite{massel_microwave_2011}. Here we consider the case where one mechanical mode, say with eigenfrequency $\omega_m$, corresponds to the cavity mode, and another one, $\omega_n$, to the mechanical mode. We also assume that $\omega_m > 2\omega_n$.  Let us discuss what happens if the system is driven with a frequency $\omega_d = \omega_m-\Delta$, $\Delta \approx \pm \omega_n$ and probed around $\omega_n$.

\begin{figure}
 \includegraphics[width=\columnwidth]{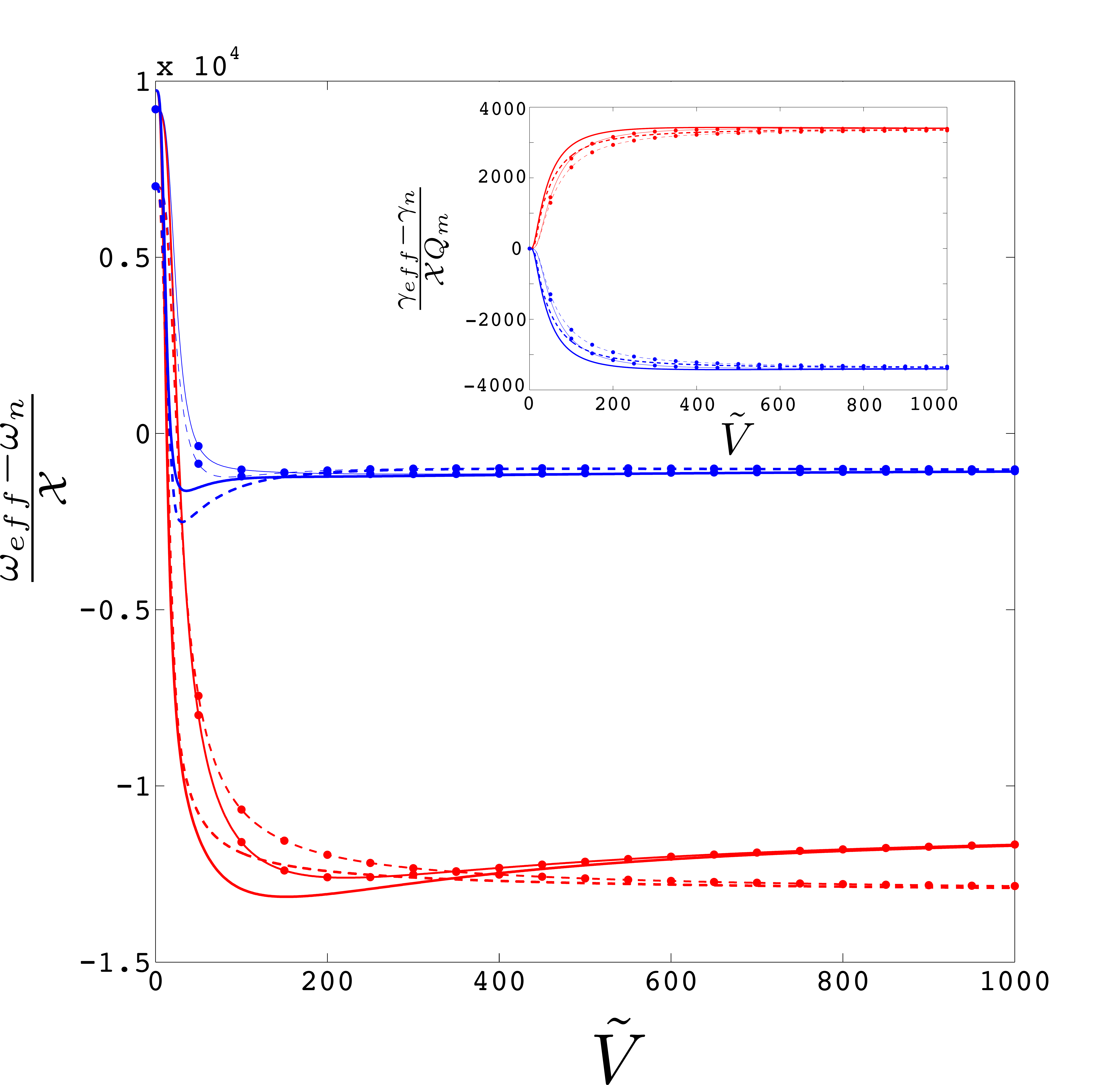}
\caption{(Color online) Effective frequency and  damping (inset) of mode $n=1$ when driving mode $m=3$ with initial tension $\tau_0=0$ (no symbols) and $\tau=10$ (circles)  in the case of red detuning (red lines, lower) with $\Delta=\omega_1$ and  in the case of blue detuning (blue lines, upper) with $\Delta=-\omega_1$. The full lines are numerical results obtained by solving the full Euler--Bernoulli equation obtained by requiring $u(x,t)$ to minimize the energy in \eqref{fullenergy}, and dashed lines  follow Eqs.~\eqref{eq:tensionequation}, \eqref{springeffect} and \eqref{gammaefft}. Here $\chi=4m\omega_s^2|\alpha_m|^2x_{zp_m}^2\frac{x_{zp_n}^2}{h^2}$ and $E_\mathrm{stress}=ESh^4/(8L^3)$.}
\label{fig:inittensionp}
\end{figure}

Neglecting other modes, the Hamiltonian is of the form
\begin{multline}
H=\omega_n \hat a_n^\dagger \hat a_n + \omega_m \hat a_m^\dagger a_m + T_n \hat x_n^3+ T_m \hat x_m^3 \\
 + T_{nnm}\hat x_n^2 \hat x_m + T_{nmm} \hat x_n \hat x_m^2+F_n \hat x_n^4 + F_m \hat x_m^4 \\
+ F_{nnnm}\hat x_n^3 \hat x_m + F_{nnmm} \hat x_n^2 \hat x_m^2+F_{nmmm}\hat x_n \hat x_m^3,
\label{eq:coupledham}
\end{multline}
where $T_{nmo}$ and $F_{nmop}$ are the sum of all the permutations of  indices $n$, $m$, $o$, $p$ of $\mathbb{T}_{nmo}$ and $\mathbb{F}_{nmop}$, respectively, and  $\mathbb{T}_{nmo}=\frac{m\omega_s^2}{h}x_{zp_n}x_{zp_m}x_{zp_o}\left(\Lambda^n_{mo}\right)_d$, $\mathbb{F}_{nmol}=\frac{m\omega_s^2}{h^2}x_{zp_n}x_{zp_m}x_{zp_l}x_{zp_o}\left(\Theta^{no}_{ml}\right)_d$, $T_n \equiv T_{nnn}$ and $T_m \equiv T_{mmm}$. Using the input/output formalism \cite{PhysRevA.31.3761} the equations of motion for  operators $\hat a_n$ and $\hat a_m$ are
\begin{eqnarray}
 \dot{\hat a}_n&=&-i\bigg[\bigg(\omega_n \hat a_n +4 F_n x_n^3+3 T_n x_n^2\\
&+&4F_{nnmm}\left(\hat a_m^\dagger \hat a_m+\frac{1}{2}\right)x_n\nonumber \\
&+&2T_{nmm}\left( \hat a_m^\dagger \hat a_m+\frac{1}{2}\right)\bigg]-\frac{\gamma_n}{2}\hat a_n+\sqrt{\gamma_n}\hat a_n^{in},\label{eqan}\nonumber \\
\dot{\hat a}_m&=&-i\bigg[\Delta \hat a_m+12 F_m \hat a_m(\hat a_m^\dagger \hat a_m) \\
&+&2(T_{mmn} \hat x_n+F_{nnmm}\hat x_n^2)\hat a_m\bigg]-\frac{\gamma_m}{2}\hat a_m+\sqrt{\gamma_m}\hat a_m^{in}.\label{eqam}\nonumber
\end{eqnarray}
Here we have written operator $a_m$ in a frame rotating with frequency $\omega_d$.
We linearize \eqref{eqan} and \eqref{eqam}, rewriting the operators as a sum of a static $\alpha$ and a fluctuating part $\delta a$, 
$\hat a= \alpha+\delta \hat a$. Keeping terms which are of the order of $\left(\frac{x_{zp_n}}{h}\right)^2$ we obtain $\alpha_n+\alpha_n^*\approx\frac{-4T_{nmm}|\alpha_m|^2}{\omega_n}$. Solving for $\delta \hat a_n$ we find the frequency response function for the input signal $\delta a_n^{in}$. It is a Lorentzian function
peaked at
\begin{equation}\label{omegaeff}
\frac{\omega_{n, \rm eff}-\omega_n}{4|\alpha_m|^2} =F_{nnmm}- 6\frac{T_nT_{nmm}}{\omega_n}\mp \frac{1}{2}\frac{T_{nmm}^2}{\omega_n}
\end{equation}
and whose width is 
\begin{equation}
\frac{\gamma_{n, \rm eff}-\gamma_n}{4|\alpha_m|^2}=\pm \frac{T_{nmm}^2}{\omega_m}Q_m.
\end{equation}
Here $Q_m=\omega_m/\gamma_m$ is the quality factor of mode $m$ and we have assumed for simplicity the fully side-band resolved limit $\omega_n \gg \gamma_m$. We remark that the results are similar to those obtained in optomechanics, the only difference comes from the second term in the effective frequency which is proportional to the self-non-linearity.
As in the case of Duffing non-linearity we consider the limit of $\tau\gtrsim1$. We find that the effective frequency when driving mode $m=3$ and probing mode $n=1$ depends on the gate voltage as
\begin{multline}
\label{springeffect}
\frac{\omega_{1, \rm eff}-\omega_1}{9 \chi \pi^4}=8
-\frac{3f_1(\tau)^2\frac{\tilde{V}^4}{\Tt^2}}{g_1(\tilde V,\tau)}%\\
\mp\frac{3}{2}\frac{f_1(\tau)^2\frac{\tilde{V}^4}{\Tt^2}}{\sqrt{g_1(\tilde V,\tau)g_3(\tilde V,\tau)}},
\end{multline}
where $g_n(\tilde V,\tau)=\frac{4I_y(n\pi)^4}{Sh^2}+\Tt(n\pi)^2+f_n(\tau)^2\frac{\tilde{V}^4}{8\Tt^2}$ and $\chi=4m\omega_s^2|\alpha_m|^2x_{zp_m}^2\frac{x_{zp_n}^2}{h^2}$ describes the amplitude of the pump. The effective damping changes as
\begin{equation}
  \frac{\gamma_{1, \rm eff}-\gamma_1}{\chi Q_3}=\pm \frac{2(3\pi)^4 f_1(\tau)^4\frac{\tilde{V}^4}{4\Tt^2}}{g_3(\tilde V,\tau)} \label{gammaefft}.
\end{equation}
In Fig.~\ref{fig:inittensionp} we plot the effective frequency  and the effective damping when driving mode $m=3$ and probing its effect on mode $n=1$. We see that both for the red- and blue-detuned pumping, i.e., $\Delta=\pm \omega_1$, the frequency shift induced by pumping, $\omega_{1,\rm eff}-\omega_1$,\cite{freqshiftnote} is positive at low gate voltages due to the fourth-order term $F_{1133}$ in Eq.~\eqref{omegaeff}, changes sign upon an increasing dc gate voltage, and tends to a voltage-independent value at large voltages. The fact that the overall frequency shift is in both cases negative --- in contrast to the traditional optomechanics --- results from the second term in Eq.~\eqref{omegaeff}, which reflects the effect of the self-non-linearity $T_n$, and which is independent of the sign of $\Delta$. This behavior applies only to the combination $n=1$, $m=3$. For higher-order $n$, the voltage-induced changes are small compared to the frequency shift at $V_{\rm dc}=0$. However, choosing $n=1$ and higher $m$ results into more complex behavior and the spring effect $\omega_{1,\rm eff}-\omega_1$ may change sign more than once in the case of blue detuning (see \cite{supplement}).

On the other hand, the change in the effective damping (inset of Fig.~\ref{fig:inittensionp}) depends on the sign of $\Delta$. For red detuning, $\Delta=\omega_n$, $\gamma_{n, \rm eff}$ increases as the voltage is increased, whereas for blue detuning $\gamma_{n,\rm eff}$ decreases. For a fixed amount of fluctuations coupling to mode $n$, the increase in damping leads to (side-band) cooling \cite{Teufel:2011jga}, whereas the decreasing damping leads to heating and, when $\gamma_{\rm eff}$ becomes zero, to a parametric instability \cite{kippenberg05}. Between these regimes, the blue-detuned driving can be used for signal amplification \cite{massel_microwave_2011}.

% \begin{equation}
%   \omega_{1, \rm eff}-\omega_1 \overset{\tilde V \gg \tau_0, 1}\longrightarrow 13000\chi,
% \end{equation}
% whereas the effective damping increases with an increasing gate voltage until it reaches a voltage independent value 
% \begin{equation}
%  \gamma_{1, \rm eff}-\gamma_1 \overset{\tilde V \gg \tau_0, 1}\longrightarrow 3400\chi Q_3.
% \end{equation}
% In Fig.~\ref{fig:inittensionp} we see that increasing the drive amplitude leads to an increase of the effective damping of mode $n$, $\gamma_{n, \rm eff}$. Such a change without an accompanying change in fluctuations leads to the (side-band) cooling of such a mode \cite{Teufel:2011jga}.

% For blue-detuned driving with $\Delta=-\omega_1$, the optical spring effect $\omega_{1, \rm eff}-\omega_1$ increases until it saturates at large voltages to the value 
% \begin{equation}
%  \omega_{1, \rm eff}-\omega_1 \overset{\tilde V \gg \tau_0, 1}\longrightarrow-1000\chi
% \end{equation}
%  while the damping decreases until it reaches the value
% \begin{equation}
% \gamma_{1, \rm eff}-\gamma_1=\overset{\tilde V \gg \tau_0, 1}\longrightarrow-3400\chi Q_3
% \end{equation}
% In Fig.~\ref{fig:inittensionp}, contrary to the previous case, for blue detuning increasing the drive amplitude leads to a decrease of the effective damping of mode $n$, allowing for an amplification of an incoming signal with frequency $\omega_n$ 

In conclusion we have derived the Hamiltonian of a thin doubly clamped nanomechanical resonator taking into account the
non-linearities between the amplitudes of the flexural modes induced by a nearby gate voltage. Besides the Duffing
non-linearity, we also find a third order non-linearity directly related to the DC deformation of the beam. This third
order non-linearity adds to the Duffing non-linearity and changes the behavior of the frequency response function. Besides
the self-non-linearity described by the Duffing behavior, we find that the different modes of the beam are non-linearly
coupled. The effective Hamiltonian of a pair of such modes resembles that of a mechanical degree of freedom coupled to a
cavity, with the difference that in the current setup the cavity is replaced by another flexural mode. Therefore, such a
coupling offers the possibility of observing the motion of one mode by observing its effect on another mode. Such
effects are the spring effect and the changing damping, and the latter can be used for side-band cooling or
amplification of a given mechanical mode.

Besides the non-linearity induced by bending described here, there may
be other sources of non-linearity in thin metallic beams, such as those
related with non-linearities in electronic properties
\cite{castellanos-gomez12} or non-linearities induced by
stretching. Our results help to identify the direct bending-induced
non-linearities and therefore facilitate  the precise tuning of
nanomechanical resonances.

We thank Mika Sillanp\"a\"a, Sung Un Cho and Xuefeng Song for useful
discussions. This work is supported in part by the Academy of Finland
and by the European Research Council (Grant No. 240362-Heattronics).

%\bibliography{./modecoupling} 

\begin{thebibliography}{10}
\bibitem{quantumnote}%
  \BibitemOpen
  \bibinfo {note} {Even though in the following we concentrate on driving
  amplitudes containing many photons, and therefore deal with essentially
  classical non-linearities, also quantum effects related with zero-point
  motion could be discussed with the Hamiltonian we derive.}%
  \bibAnnoteFile{Stop}{quantumnote}%
\bibitem{freqshiftnote}%
  \BibitemOpen
  \bibinfo {note} {Note that tuning the gate voltage also changes the bare
  frequencies $\omega_n$. What we consider here is the additional frequency
  shift due to the driving of mode $m\neq n$.}%
  \bibAnnoteFile{Stop}{freqshiftnote}%
\end{thebibliography}

\bibliographystyle{}

\begin{widetext}
\newpage
\section*{Appendix : Derivation of the non-linear Hamiltonian}\label{appendix}
Expanding $u_{ac}$ in Eq.~(2) of the main text on an arbitrary basis $\chi_n(x)$, $u_{ac}=h \sum_n y_n\chi_n$ we get the Hamiltonian of the form
\begin{equation}\label{energywithnonlinear}
H=\frac{p_n^2}{2m}+E_{\rm stress}\left\{\left[\Omega^2 \right]^n_m y_ny^m+\Lambda^n_{mo}y_ny^my^o
+\Theta^{no}_{mp}y_ny^my_oy^p\right\}+E_g,
\end{equation} 
with the non-linear coefficients
\begin{subequations}
\label{eq:hamiltoniancoefficients}
\begin{eqnarray}
\left[\Omega^2 \right]^n_m &=&\frac{\omega_0^2}{\omega_s^2}\ddcnddcm+\tau \dcndcm
+2\dcndcm\dudu+4\dudcn\dudcm\nonumber \\ &-&
\frac{V_{dc}^2}{m\omega_s^2h^2}\int^1_0\frac{d^2C[d-u(x,t)]}{du_{ac}^2}\cncm dx \\
\Lambda^n_{mo}&=&4\dudco\dcndcm \label{tco}-\frac{V_{dc}^2}{m\omega_s^2h^2}\int_0^1\frac{d^3C[d-u(x,t)]}{du_{ac}^3}\chi_n\chi_m\chi_odx \label{tud}\\
\Theta^{no}_{mp}&=&\dcndcm\dcodcp -\frac{V_{dc}^2}{m\omega_s^2h^2}\int_0^1\frac{d^4C[d-u(x,t)]}{du_{ac}^4}\chi_n\chi_m\chi_p\chi_odx\bigg)\\
E_g&=&(2V_{dc}V_{ac}+V_{ac}^2)\int_0^1f(u_0(x))dx.
\end{eqnarray}
\end{subequations}
\end{widetext}
Here $C[d]$ is the capacitance between the gate and the beam at the distance $d$, $\frac{1}{2}m \omega_0^2h^2=\frac{EI}{2L^3}$ is the bending energy of a beam displaced by $h$ and $\frac{1}{2}m \omega_s^2h^2=\frac{ESh^4}{8L^3}$ is the stress energy of the beam displaced by $h$ with respect to its equilibrium position. The coefficient $E_g$ describes the feedback of the motion of the resonator on the gate voltage and is neglected below as we assume a fixed voltage drive. Besides the voltage, the system is described by the two dimensionless parameters $\tau_0=4T_0L^2 /(E S h^2)$ and $\left(\omega_0/\omega_s\right)^2=4I_y/( S h^2)$. For a rectangular beam,  which we consider in the following, $\left(\omega_0/\omega_s\right)^2=1/3$. Overall, our main results do not greatly depend on $\omega_0$. The thickness $h$ appears in the above expressions only because it sets the magnitude of the deformation --- it scales out from the final results of observable quantities. 

 Writing the Hamiltonian in the basis  which diagonalises $\Omega$ and scaling $h y_n$ by the amplitude of the zero-point motion $\sqrt{1/m \omega_n}$ 
one arrives at Eq.~(1) of the main text. We consider here specifically  a doubly clamped beam. For low voltage the DC deformation $u_{dc}$ is given by the Euler--Bernoulli equation in the case of a parallel plate capacitance model:
\begin{equation}\label{EB}
m\omega_0^2h\, u_{dc}''''-\left(m\omega_T^2h+2m\omega_s^2h\int_0^1u_{dc}'^2dx\right)u_{dc}''=\frac{V_g^2}{2}\frac{\epsilon W L}{d^2}.
\end{equation}
The solution of this integro-differential equation is \cite{PhysRevB.67.235414}
\begin{equation}
\label{eq:udc}
\begin{split}
 u_{dc}=\frac{\tilde{V}^2}{8\tau\xi}\bigg(&\coth\left(\frac{\xi}{2}\right)(\cosh(\xi x)-1)-\sinh (\xi x)\\&+\xi x-\xi x^2\bigg),
\end{split}
\end{equation}
where $\tilde{V}^2=(-h \partial_d C V_{g}^2)/E_{\rm stress}$, $E_{\rm stress}=m\omega_s^2 h^2/2=ESh^4/(8 L^3)$. 
 We define the total tension
\begin{equation}
\tau=\tau_0+2\int_0^1u_{dc}'^2dx
\label{eq:totaltension}
\end{equation}
and $\xi=\sqrt{3\tau}$. Substituting Eq.~\eqref{eq:udc} into Eq.~\eqref{eq:totaltension}, integrating and then disregarding exponential terms $\sim \exp(-\xi)$ yields a self-consistency equation for $\tau$, 
\begin{equation}
  \tau=\tau_0+\frac{\tilde{V}^4}{96\tau^2}\left(1-\frac{3\sqrt{3}}{\sqrt{\tau}}+\frac{8}{\tau}\right),
\label{eq:totaltension2}
\end{equation}

% \begin{equation}
%  u_{dc}=\left(\frac{\tilde{V}}{2}\right)^2\frac{1}{\tau}\left( x- x^2\right).
% \label{eq:deformationapproximation}
% \end{equation}
\begin{widetext}
 Since in the limit of large $\tau$ Eq.~\eqref{EB} reduces to the wave equation, we  use the harmonic wave function $\chi_n=\sqrt{2}\sin(n\pi x)$ as a basis for \eqref{EB} and get 
\begin{subequations}
\begin{align}
\left[\Omega^2 \right]^n_m&=\frac{(n\pi)^4}{3}+\tau(n\pi)^2+\frac{\tilde V^4}{8\tau^2}f_n(\tau)f_m(\tau)-\frac{\tilde{V}^2}{2}\frac{h}{d}\label{OMEGA}\\ 
 \label{LAMBDA} \Lambda^n_{mo}&=\sqrt{2}(n\pi)^2 \frac{\tilde{V}^2}{2\tau}f_o(\tau)\delta^n_m-\frac{\tilde{V}^2}{2}\left(\frac{h}{d}\right)^2\frac{4 \sqrt{2} m n o \left((-1)^{m+n+o}-1\right)}{\pi  (m-n-o) (m+n-o) (m-n+o) (m+n+o)}\\
\Theta^{no}_{mp}&=(n\pi)^2(p\pi)^2\delta^n_{m}\delta^o_{p}\label{THETA}-\frac{\tilde{V}^2}{2}\left(\frac{h}{d}\right)^3\frac{1}{2}[\delta_{n+m+o,-p}+\delta_{n+m,o+p}+\delta_{n+p,m+o}+\delta_{n+o,m+p}-\delta_{n+m+o,p}\\
&\quad \quad \quad \quad \quad \quad \quad \quad \quad \quad \quad \quad \quad \quad \quad-\delta_{n+m+p,o}-\delta_{n+o+p,m}-\delta_{m+o+p,n}] \nonumber\\
f_n(\tau)&=n\pi  \left(1-(-1)^n\right)  \left(\frac{2}{\pi ^2 n^2}-\frac{ \sqrt{3\tau} }{\pi ^2 n^2+3 \tau}\right).
\label{fn}
\end{align}
\end{subequations}
\end{widetext}
The last terms in Eqs.~\eqref{OMEGA}, \eqref{LAMBDA} and \eqref{THETA} are relevant only for low voltages in the presence of initial tension, and do not greatly contribute to the physics discussed in this paper. We thus drop them out in the ensuing analytical approximations. However in the numerical results, we use the full solutions of the Euler--Bernoulli equation to determine the eigenmodes and the coupling constants. Nevertheless, Eqs.~(\ref{OMEGA}-\ref{THETA}) represent a fairly approximations in the limit of relatively strong tension.

\begin{figure}[h]
 \includegraphics[width=\columnwidth]{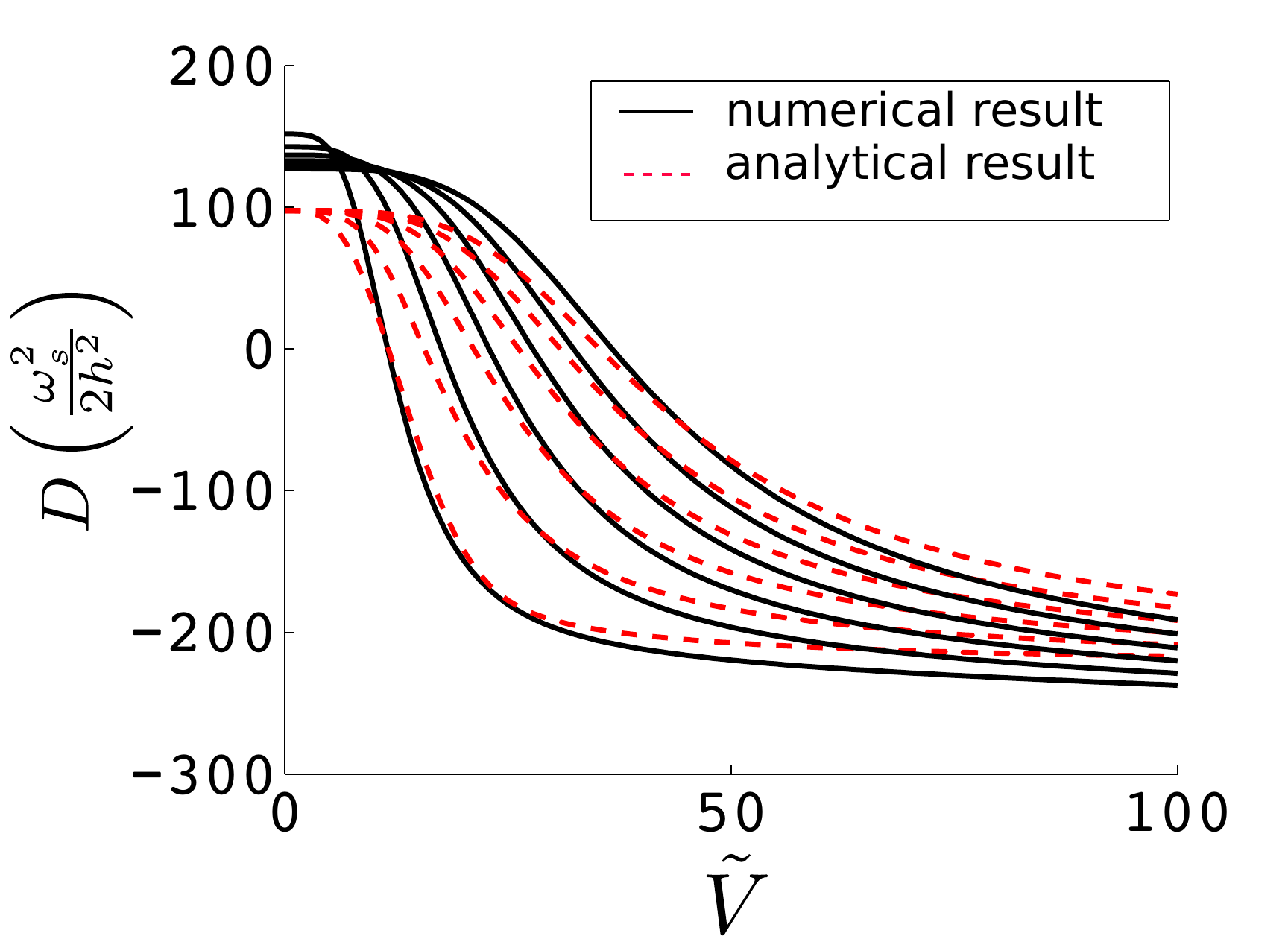}
\caption{Duffing  constant for the first mode with different initial tension $ \tau$. From left to right the dimensionless initial tension  $\tau_0$ starts from 0 and increases with the step of 10. The dashed lines are our analytic expressions and the full line are numerical solutions of the full Euler--Bernoulli equations obtained from Eq.~(2) of the main text. The deviation between the two set of curves at low $V$ is due to our scheme of approximating mode functions by harmonic functions. }
\label{fig:duffingvsvoltage}
\end{figure}

\begin{figure}[h]
 \includegraphics[width=\columnwidth]{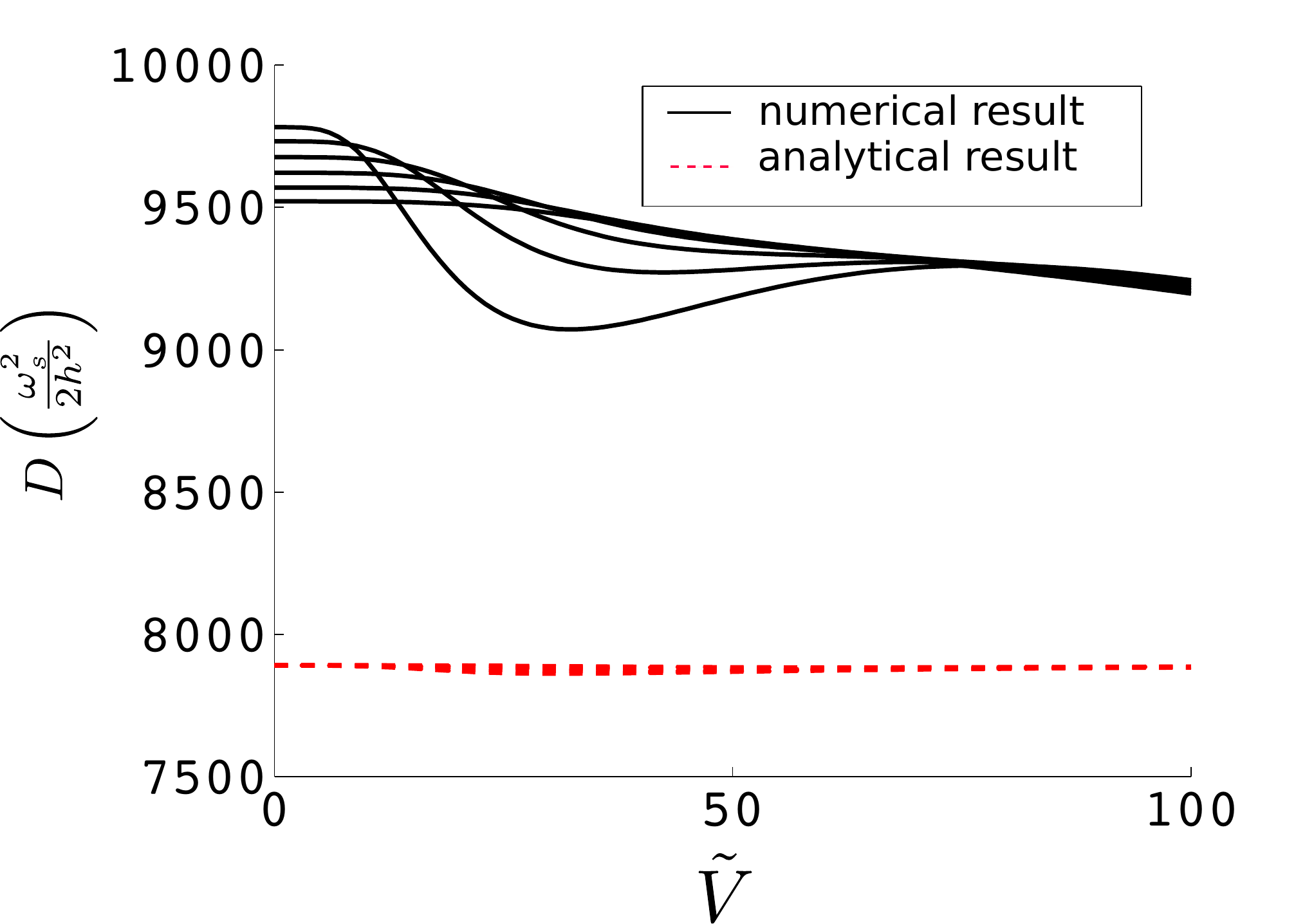}
\caption{Duffing  constant for the third mode with different initial tensions $ \tau_0$. From left to right the dimensionless initial tension  $\tau_0$ starts from 0 and increases with the step of 10. The dashed lines are our analytic expressions and the full line are numerical solutions of the full Euler--Bernoulli equations obtained from Eq.~(2) of the main text. The approximation with harmonic mode functions underestimates $\Theta_{mp}^{no}$, which is the reason for the discrepancy between the full numerical solutions and our analytic approximations.}
\label{fig:duffingvsvoltage3}
\end{figure}

\section*{Duffing constant}

\begin{figure}
 \includegraphics[width=\columnwidth]{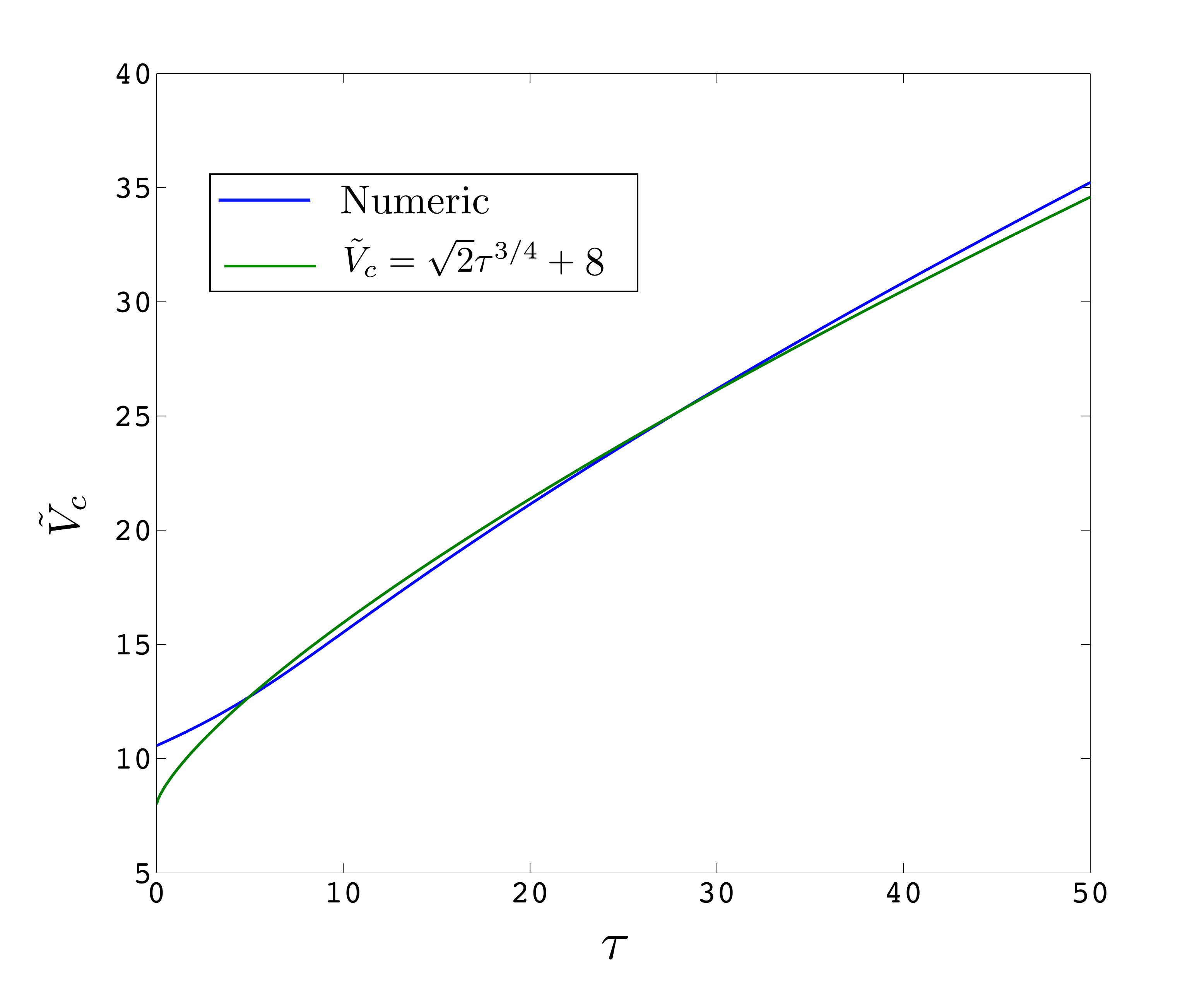}
\caption{Crossover voltage $V_c$ for the sign change of the Duffing constant with respect to the initial tension $\tau$. }
\label{fig:vcvctau}
\end{figure}

We can estimate the overall behavior of the Duffing constant $D=\omega_s^2/(2h^2)\left[(\Theta_{nn}^{nn})_d-\frac{5}{4}(\Lambda_{nn}^n)_d^2/[2(\Omega_n^n)_d^2]\right]$ by disregarding the off-diagonal terms in Eqs.~(\ref{OMEGA}-\ref{LAMBDA}) and solving the tension $\tau$ with Eq.~\eqref{eq:totaltension2}.
The tension $\tau$ exhibits a rather complicated voltage dependence, however its behaviour can be investigated in different  limiting cases. A characteristic value for the voltage can be found by substituting  $\tau\equiv \tau_0$ into Eq.~\eqref{eq:totaltension2} and comparing $\tau_0$ to the second  terms of the right hand side of Eq.~\eqref{eq:totaltension2}. This yields
 \begin{equation}
  \tilde{V}^*=96^{1/4}\frac{\tau_0^{3/4}}{\left(1-3\frac{\sqrt{3}}{\sqrt{\tau_0}}+\frac{8}{\tau_0}\right)^{1/4}}.
 \end{equation}
Thus we find that for $\tilde{V}\ll\tilde{V}^*$ the tension $\tau\approx \tau_0$ while for $\tilde{V}\gg\tilde{V}^*$ we have $\tau\approx \frac{V^{4/3}}{96^{1/3}}$. 
Disregarding terms coming from the electrostatic force  the general expression for the Duffing constant is
\begin{equation}
D=\frac{\omega_s^2}{2h^2}\left[(n\pi)^4-\frac{5}{8}\frac{(n\pi)^4f_n(\tau)^2\frac{V^4}{\tau^2}}{\frac{4I_y}{Sh^2}(n\pi^4)+\tau(n\pi)^2+f_n(\tau)^2\frac{V^4}{8\tau^2}}\right].
\end{equation}
As shown in Figs.~\ref{fig:duffingvsvoltage}, \ref{fig:duffingvsvoltage3},  at low $\tilde V $, $D$ starts from a positive value $D(V=0)\approx  \omega_s^2/(2h)^2(n\pi)^4$  and tends to a voltage-independent value $D(\tilde V ) = \omega_s^2/(2h^2)\left[\frac{(n\pi)^4[(n\pi)^4-768)}{192+(n\pi)^4}\right]$ at large voltages.
Note that, for a symmetric dc deformation, this behavior is only valid for odd-order modes (with symmetric eigenfunctions with respect to the center of the beam). Indeed, from Eq.~\eqref{fn}, we find that the Duffing constant is voltage independent for an even $n$.
In Fig.~\ref{fig:duffingvsvoltage}, we plot the behaviour of the Duffing constant for the first mode for different values of $\tau$. We find that $D$ changes its sign for a given value of the voltage, $\tilde{V}=\tilde{V}_c$ and the effect of the initial tension is to shift the crossover voltage to higher values (Fig.~\ref{fig:vcvctau}). For $n>1$ the Duffing constant varies only weakly with $\tilde{V}$ and stays always positive (see Fig.~\ref{fig:duffingvsvoltage3}.

\section*{Mode coupling}

The frequency response function $f$ for the input signal $\delta a_n^{in}$ solved from Eqs.~(10-11) of the main text is
\begin{widetext}
 \begin{equation}
 f=-i \sqrt{\gamma_n}\frac{\left[\Delta^2 +\left(\frac{\gamma_m}{2}-i \omega\right)^2-c^2|\alpha_m|^4\right] \left[a+i \frac{\gamma_n}{2}+\omega+\omega_n\right]-2 b^2 |\alpha_m|^2(\Delta+c|\alpha_m|^2)}{\left[\Delta^2 +\left(\frac{\gamma_m}{2}-i \omega\right)^2-c^2|\alpha_m|^4\right]\left[\left(\frac{\gamma_n}{2}-i\omega\right)^2+\omega_n ^2+2a\omega_n\right]-4\omega_1 b^2 |\alpha_m|^2(\Delta+c|\alpha_m|^2)}
\end{equation}
%l
%\end{widetext}
with 
\begin{eqnarray*}
 a&=&12 F_n (\alpha_n+\alpha_n^*)^2+6 T_n(\alpha_n+\alpha_n^*)+4F_{nnmm}|\alpha_m|^2+2F_{nnmm}\\
b&=&4F_{nnmm}(\alpha_n+\alpha_n^*)+2T_{nmm}\\
%c&=&24 F_3|\alpha_m|^2+2T_{nmm}(\alpha_n+\alpha_n^*)+2 F_n (\alpha_n+\alpha_n^*)^2\\
c&=&12 F_m.
\end{eqnarray*}
This frequency response function describes a Lorentzian resonance around an effective frequency $\omega_{n, \rm eff}$ with damping $\gamma_{n, \rm eff}$ described in Eqs.~(12) and (13) of the main text. Using the approximations leading to Eqs.~\eqref{eq:hamiltoniancoefficients} we find the effective frequency
% Keeping terms which are of the order of $\left(\frac{x_{zp_n}}{h}\right)^2$ we find $\alpha_n+\alpha_n^*\approx\frac{-4T_{133}|\alpha_3|^2}{\omega_n}$. Therefore we find that for $\Delta\approx \pm \omega_n$ the effective frequency and the effective damping are
% \begin{equation}\label{omegaeff}
% \frac{\omega_{n, \rm eff}-\omega_n}{4|\alpha_m|^2} =F_{nnmm}- 6\frac{T_nT_{nmm}}{\omega_n}\mp \frac{1}{2}\frac{T_{nmm}^2}{\omega_n}
% \end{equation}
% \begin{equation}
% \frac{\gamma_{n, \rm eff}-\gamma_n}{4|\alpha_m|^2}=\pm \frac{T_{nmm}^2}{\omega_m}Q_m.
% \end{equation}
% Here $Q_m$ is the quality factor of  mode $m$. We remark that the results are similar to those obtained in optomechanics, the only difference comes from the second term in the effective frequency which is proportional to the self nonlinearity.
% As in the case of Duffing nonlinearity we consider the limit of $\tau\gtrsim1$. We find that the general expression for the effective frequency is 
%\begin{widetext}
\begin{eqnarray*}
\frac{\omega_{n, \rm eff}-\omega_n}{\chi}&=&n^2 m^2 \pi^4\Bigg[8
-\frac{3f_n(\tau)^2\frac{\tilde{V}^4}{\Tt^2}}{\frac{4I_y(n \pi)^4}{Sh^2}+\Tt(n\pi)^2+f_n(\tau)^2\frac{\tilde{V}^4}{8\Tt^2}}\\
&\mp&\frac{m^2}{4n^2}\frac{f_n(\tau)^2\frac{\tilde{V}^4}{\Tt^2}}{\sqrt{\left(\frac{4I_y(n \pi)^4}{Sh^2}+\Tt(n\pi)^2+f_n(\tau)^2\frac{\tilde{V}^4}{8\Tt^2}\right)\left(\frac{4I_y(m \pi)^4}{Sh^2}+\Tt(m\pi)^2+f_m(\tau)^2\frac{\tilde{V}^4}{8\Tt^2}\right)}}\Bigg]
\end{eqnarray*}
%\end{widetext}
and effective damping
\begin{equation}
  \frac{\gamma_{n, \rm eff}-\gamma_n}{\chi Q_m}=\pm \frac{2(m\pi)^4f_n(\tau)^2\frac{\tilde{V}^4}{4\Tt^2}}{\frac{4I_y(m \pi)^4}{Sh^2}+\Tt(m\pi)^2+f_m(\tau)^2\frac{\tilde{V}^4}{8\Tt^2}}\label{gammaefft}.
\end{equation}
Here $\chi=4m\omega_s^2|\alpha_m|^2x_{zp_m}^2\frac{x_{zp_n}^2}{fh^2}$ describes the amplitude of the pump. 

In Fig.~3 of the main text, we plot the effective frequency  and the effective damping when driving mode $m=3$ and probing its effect on mode $n=1$. We see that, for red-detuned pumping with $\Delta=\omega_n$, the frequency shift induced by pumping is positive at low gate voltages, $\omega_{n, \rm eff}(V_g=0)-\omega_n=(n\pi)^2(m\pi)^2\chi$, changes sign with increasing dc gate voltage, and tends to a voltage-independent value at large voltages
%\begin{widetext}
\begin{equation}
  \omega_{n, \rm eff}-\omega_n \overset{\tilde V \gg \tau_0, 1}\longrightarrow \chi(nm)^2\pi^4\left[8-\frac{4608/(n\pi)^2}{(n\pi)^2+192/(n\pi^2)}-\frac{m^2}{n^2}\frac{384/(n\pi)^2}{\sqrt{[(n\pi)^2+192/(n\pi^2)][(m\pi)^2+192/(m\pi^2)]}}\right].
\end{equation}
%\end{widetext}
On the other hand the effective damping increases with an increasing gate voltage until it reaches a voltage-independent value 
\begin{equation}
 \frac{\gamma_{n, \rm eff}-\gamma}{\chi Q_m}=\frac{m^6}{n^4}\frac{768 \pi ^4}{\pi ^4 m^4+192}.
\end{equation}
For blue-detuned driving, when $\Delta=-\omega_n$, the optical spring effect $(\omega_{n, \rm eff}-\omega)$ increases until it saturates at large voltages to the value 
%\begin{widetext}
\begin{equation}
  \omega_{n, \rm eff}-\omega_n \overset{\tilde V \gg \tau_0, 1}\longrightarrow \chi(nm)^2\pi^4\left[8-\frac{4608/(n\pi)^2}{(n\pi)^2+192/(n\pi^2)}+\frac{m^2}{n^2}\frac{384/(n\pi)^2}{\sqrt{[(n\pi)^2+192/(n\pi^2)][(m\pi)^2+192/(m\pi^2)]}}\right]
\end{equation}
\end{widetext}

\begin{figure}[h!]
\includegraphics[width=\columnwidth]{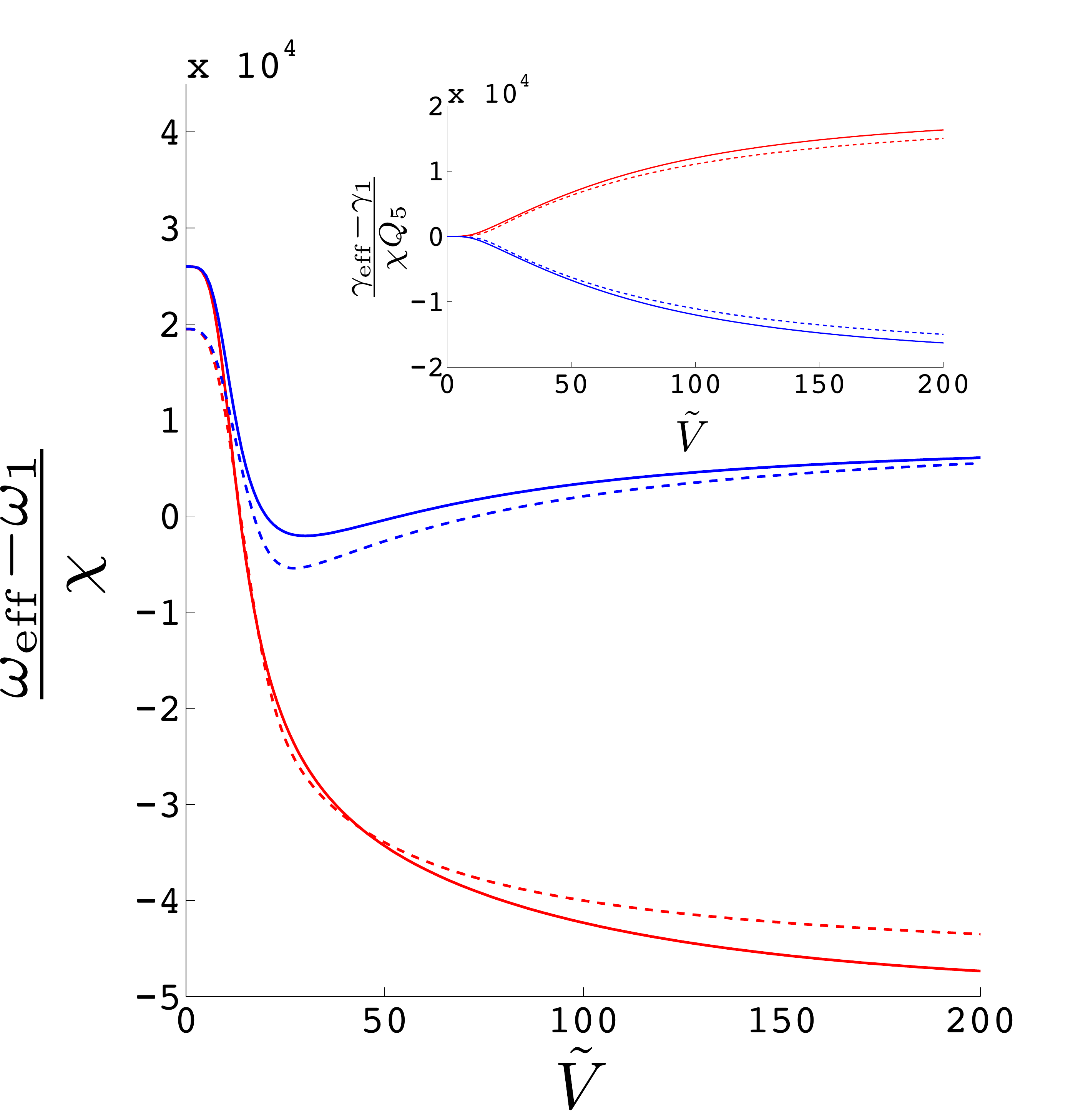}
\caption{Effective frequency and damping (inset) of mode n = 1 when driving mode
m = 5 and when initial tension $\tau_0$ = 0  in the case of red detuning (red lines, lower) with
$\Delta = \omega_1$ and in the case of blue detuning (blue lines, upper)
with $\Delta = -\omega_1$. The full lines are numerical results obtained
by solving the full Euler--Bernoulli equation and dashed lines
are analytical results derived in the text.
}
\label{fig:omega15}
\end{figure}
\begin{figure}[h!]
\includegraphics[width=\columnwidth]{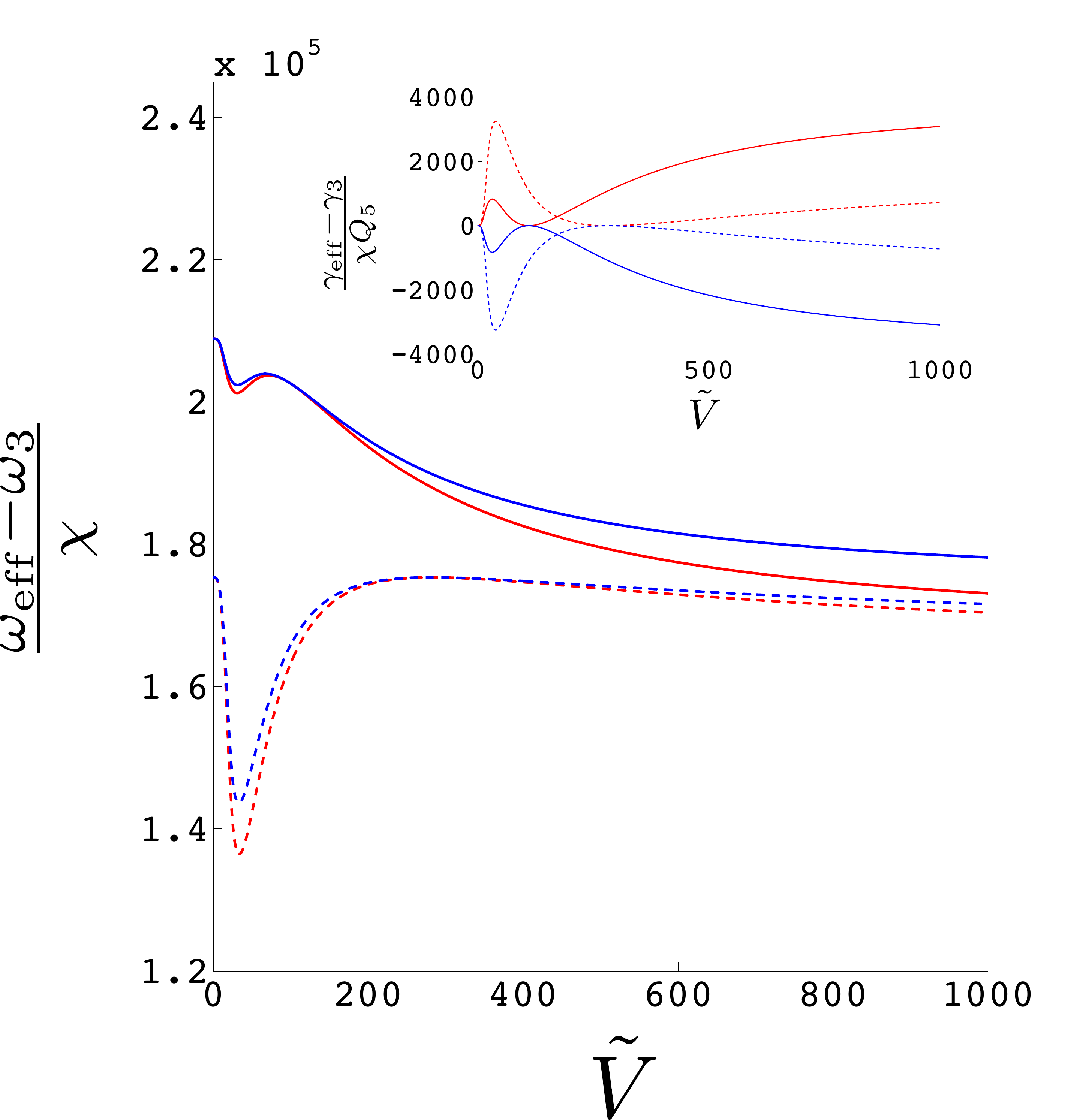}
\caption{Effective frequency and damping (inset) of mode n = 3 when driving mode
m = 5 and when initial tension $\tau_0$ = 0  in the case of red detuning (red lines, lower) with
$\Delta = \omega_3$ and in the case of blue detuning (blue lines, upper)
with $\Delta =- \omega_3$. The full lines are numerical results obtained
by solving the full Euler--Bernoulli equation and dashed lines
are analytical results are analytical results derived in the text.
}
\label{fig:omega35}
\end{figure}
while the damping decreases until it reaches the value
\begin{equation}
\frac{\gamma_{n, \rm eff}-\gamma}{\chi Q_m}=-\frac{m^6}{n^4}\frac{768 \pi ^4}{\pi ^4 m^4+192}
\end{equation}
These predictions are compared to the full numerical solutions obtained from the Hamiltonian Eq.~(2) of the main text in Fig.~3 of the main text.
Although we are focusing only on the first and third mode in the main text one can also consider pairs of other modes. We plot the effective frequency and effective damping for $n=1$ and $m=5$ in Fig.~\ref{fig:omega15} and for $n=3$ and $m=5$ in Fig.~\ref{fig:omega35}. In the first case  when driving mode $m=5$ and measuring mode $n=1$ the change in the effective frequency is larger than the one we have when driving mode $m=3$. We also find that in the case of blue detuning the 
effective frequency changes its sign twice as a function of the gate voltage. When driving mode $m=5$ and measuring mode $n=3$, the voltage dependence of the change in the effective frequency is relatively small as the fourth-order term dominates throughout the interesting regime of voltages.

%  \begin{figure}

\end{document}